\documentclass[ss,preprint]{imsart}

\RequirePackage[OT1]{fontenc}
\RequirePackage{amsthm,amsmath}
\RequirePackage[numbers]{natbib}
\RequirePackage[colorlinks,citecolor=blue,urlcolor=blue]{hyperref}

\usepackage{graphicx}
\pubyear{2006}
\volume{0}
\issue{0}
\firstpage{1}
\lastpage{8}

\startlocaldefs
\numberwithin{equation}{section}
\theoremstyle{plain}

\endlocaldefs

\begin{document}

\begin{frontmatter}
\title{Two samples test for discrete power-law distributions}
\runtitle{Two samples test for discrete power-law distributions}

\begin{aug}
\author{\fnms{Alessandro} \snm{Bessi}\ead[label=e1]{alessandro.bessi@iusspavia.it}}

\address{IUSS Institute for Advanced Study, Pavia, Italy\\
\printead{e1}}

\end{aug}

\begin{abstract}
Power-law distributions occur in wide variety of physical, biological, and social phenomena. In this paper, we propose a statistical hypothesis test based on the log-likelihood ratio to assess whether two samples of discrete data are drawn from the same power-law distribution.
\end{abstract}

\begin{keyword}
\kwd{power-law} 
\kwd{log-likelihood ratio}
\kwd{statistical hypothesis test}
\end{keyword}
\tableofcontents
\end{frontmatter}

\section{Introduction}
Power-law distributions have attracted particular attention for their mathematical properties and appearances in a wide variety of scientific contexts, from physical and biological sciences to social and man-made phenomena. 

Differently from those Normally distributed, empirical quantities distributed according to a power-law do not cluster around an average value, and thus we can not characterize a power-law distribution through its mean and standard deviation. However, the fact that some scientific observations can not be characterized as simply as other measurements is often a sign of complex underlying processes that deserve further study \cite{clauset2009}.

Formally, a discrete or continuous quantity $x$ is distributed according to a power-law if its probability distribution is
$$ p(x) \propto x^{-\alpha},$$

where $\alpha$ is called the \emph{scaling parameter} of the distribution. However, only few empirical phenomena show such a  probability distribution for all values of $x$. More often, the power-law applies only for values greater than some minimum value $x_{min}$, so that only the tail of a distribution behaves according to a power-law. 

A complete introduction to power-law distributions along with a statistical framework for discerning and quantifying power-law behavior in empirical data can be found in \cite{clauset2009}, whereas extensive discussions can be found in \cite{mitzenmacher2004,newman2005,sornette2006}, and references therein.

In this paper, we limit our attention to discrete power-law distributions. In particular, we focus on the comparison of two samples of discrete data to assess whether they are drawn from the same power-law distribution. For example, we could be interested in assessing whether the number of likes received by posts on Facebook and the number of likes received by videos on YouTube follow the same power-law distribution, or whether the degrees of the nodes belonging to two separate clusters in a small-world network are distributed according to the same power-law distribution. To answer those questions we can not rely on the Kolmogorov-Smirov test \cite{conover1971,marsaglia2003}. Since such a statistical test is formulated for continuous distributions, its application to discrete distributions will produce approximate p-values due to the presence of ties.

In order to overcome such an issue, we introduce a statistical hypothesis test based on the log-likelihood ratio to assess whether two samples of discrete data are drawn from the same population. Under the null hypothesis, i.e. the two samples are drawn from the same power-law distribution, the resulting test statistics follows the $\chi^2$ distribution with $1$ degree of freedom.

This paper is structured as follows. In Section 2 we provide some basic definitions about discrete power-law distributions. In Section 3 we first introduce and discuss the proposed statistical test, and then we conclude with two examples illustrating how the test performs.

\section{Definitions}
\paragraph{Discrete power-law distributions.} Power-law distributions can be continuous or discrete. Here we focus on the discrete case, i.e. when the quantity of interest can assume only positive integers.
Let $x$ represents the quantity whose distribution we are interested in. The probability distribution is 
$$ p(x) = Pr(X = x) = \frac{x^{-\alpha}}{\zeta(\alpha,x_{min})}, $$

where 
$$\zeta(\alpha,x_{min}) = \sum_{n=0}^{\infty}(n + x_{min})^{-\alpha}$$ 

is the generalized or Hurwitz zeta function. It is useful to consider also the complementary cumulative distribution function,
$$ P(x) = Pr(X > x) = \frac{\zeta(\alpha,x)}{\zeta(\alpha,x_{min})},$$

since it allows to plot power-law distributions in doubly logarithmic axes, and thus emphasize the upper tail behavior.

\paragraph{Estimating the scaling parameter.} The method for fitting parametrized models to observed data is the method of maximum likelihood, which provably gives accurate parameter estimates in the limit of large sample size \cite{cox1995,wasserman2003}. The likelihood function of a power-law distribution given a sample $x = (x_{1},x_{2},\dots,x_{n})$ is
$$ L(\alpha,x_{min}) = \alpha^{n} x_{min}^{n\alpha} \prod_{i = 1}^{n} \frac{1}{x_{i}^{\alpha+1}},$$

whereas the log-likelihood function is
$$ l(\alpha,x_{min}) = n  \mathrm{ln} \alpha + n\alpha \mathrm{ln}x_{min} + (\alpha+1)\sum_{i=1}^{n}\mathrm{ln}x_{i}.$$

Assuming that data are drawn from a power-law distribution with $x \geq x_{min}$, we can derive a maximum likelihood estimators (MLE) of the scaling parameter $\alpha$. Although there is no exact closed form expression for the MLE in the discrete case, an approximate expression can be derived using an approach that considers power-law distributed integers approximated as continuous reals rounded to the nearest integer (details of the derivation are given in Appendix B of \cite{clauset2009}). The result is
$$\hat{\alpha} \simeq 1 + n \left[ \sum_{i = 1}^{n} \mathrm{ln}\frac{x_{i}}{x_{min} - \frac{1}{2}} \right]^{-1}.$$

\section{Two samples test for discrete power-law distributions}

\paragraph{Statistical hypothesis test.} Suppose that we have two samples of discrete data, $s_{1}$ and $s_{2}$, and we want to assess whether the two samples are drawn from the same power-law distribution. We can not rely on the Kolmogorov-Smirnov test. Indeed, since such a statistical test is formulated for continuous distributions, its application to discrete distributions will produce approximate p-values due to the presence of ties. In order to overcome such an issue, we propose a statistical test based on the log-likelihood ratio:
$$ \Lambda = -2 \times l(H_{0}|s_{1} \cup s_{2}) + 2 \times \left[ l(H_{1}|s_{1}) + l(H_{1}|s_{2}) \right],$$

where $ l(H_{0}|s_{1} \cup s_{2})$, i.e. the null model, is the log-likelihood of the pooled samples, $s_{1} \cup s_{2}$, whereas $l(H_{1}|s_{1}) + l(H_{1}|s_{2})$, i.e. the alternative model, is the sum of the log-likelihoods of the samples $s_{1}$ and $s_{2}$.

Under the null hypothesis, i.e. the two samples are drawn from the same power-law distribution, the obtained test statistics $\Lambda$ will follow a $\chi^{2}$ distribution with $1$ degree of freedom. Indeed, $df = 2 - 1 = 1$, since in the alternative hypothesis we need to estimate two parameters, $\hat{\alpha}_{s_{1}}$ and $\hat{\alpha}_{s_{2}}$, whereas in the null model we need to estimate only one parameter $\hat{\alpha}_{s_{1} \cup s_{2}}$.
 
\begin{figure}[h]
	\centering\includegraphics[width = 0.7 \textwidth]{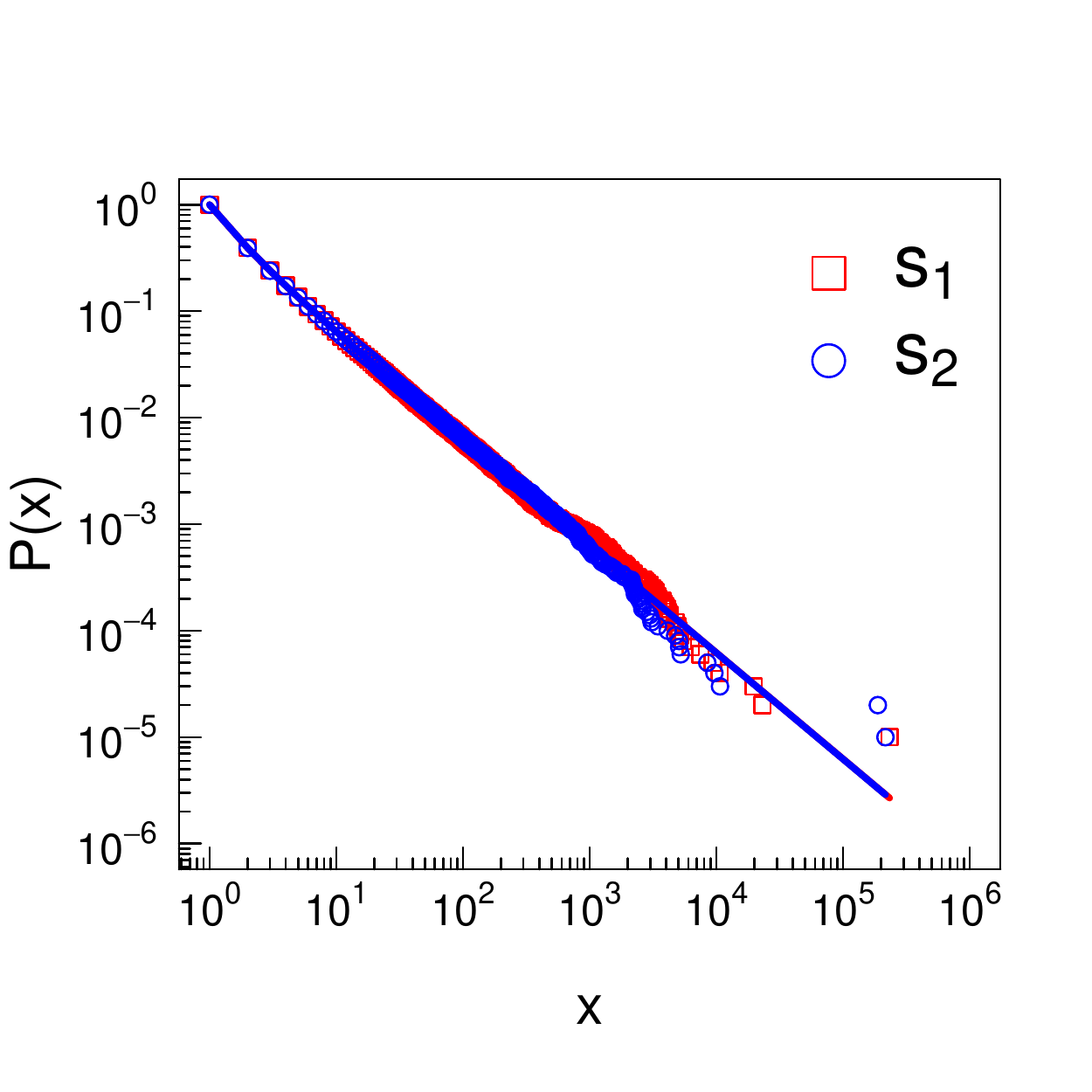}
	\caption{\textbf{Two samples drawn from the same distribution.} Two samples of discrete data $s_{1}$ ($n_{s_{1}} = 10^{5}$) and $s_{2}$ ($n_{s_{2}} = 10^{5}$) are compared. The estimates of the scaling parameters are, respectively, $\hat{\alpha}_{s_{1}} = 1.997852$ and $\hat{\alpha}_{s_{2}} = 1.99816$. The test statistics is $\Lambda = 0.006508615$ with p-value $0.9356996 > 0.05$, thus we fail to reject the null hypothesis and conclude that the two samples are drawn from the same power-law distribution.} 
	\label{fig:1}
\end{figure}

\paragraph{Example 1: two samples drawn from the same distribution.} Figure \ref{fig:1} shows the complementary cumulative distribution function, $P(x)$, of two samples, $s_{1}$ ($n_{s_{1}} = 10^{5}$) and $s_{2}$ ($n_{s_{2}} = 10^{5}$), that we want to compare in order to assess whether they are drawn from the same power-law distribution.  The estimates of the scaling parameters are, respectively, $\hat{\alpha}_{s_{1}} = 1.997852$ and $\hat{\alpha}_{s_{2}} = 1.99816$. The test statistics is $\Lambda = 0.006508615$ with p-value $0.9356996 > 0.05$, thus we fail to reject the null hypothesis and conclude that the two samples are drawn from the same power-law distribution. Indeed, these samples were randomly drawn from a power-law distribution with $x_{min} = 1$ and $\alpha = 2$.

\begin{figure}[h]
	\centering\includegraphics[width = 0.7 \textwidth]{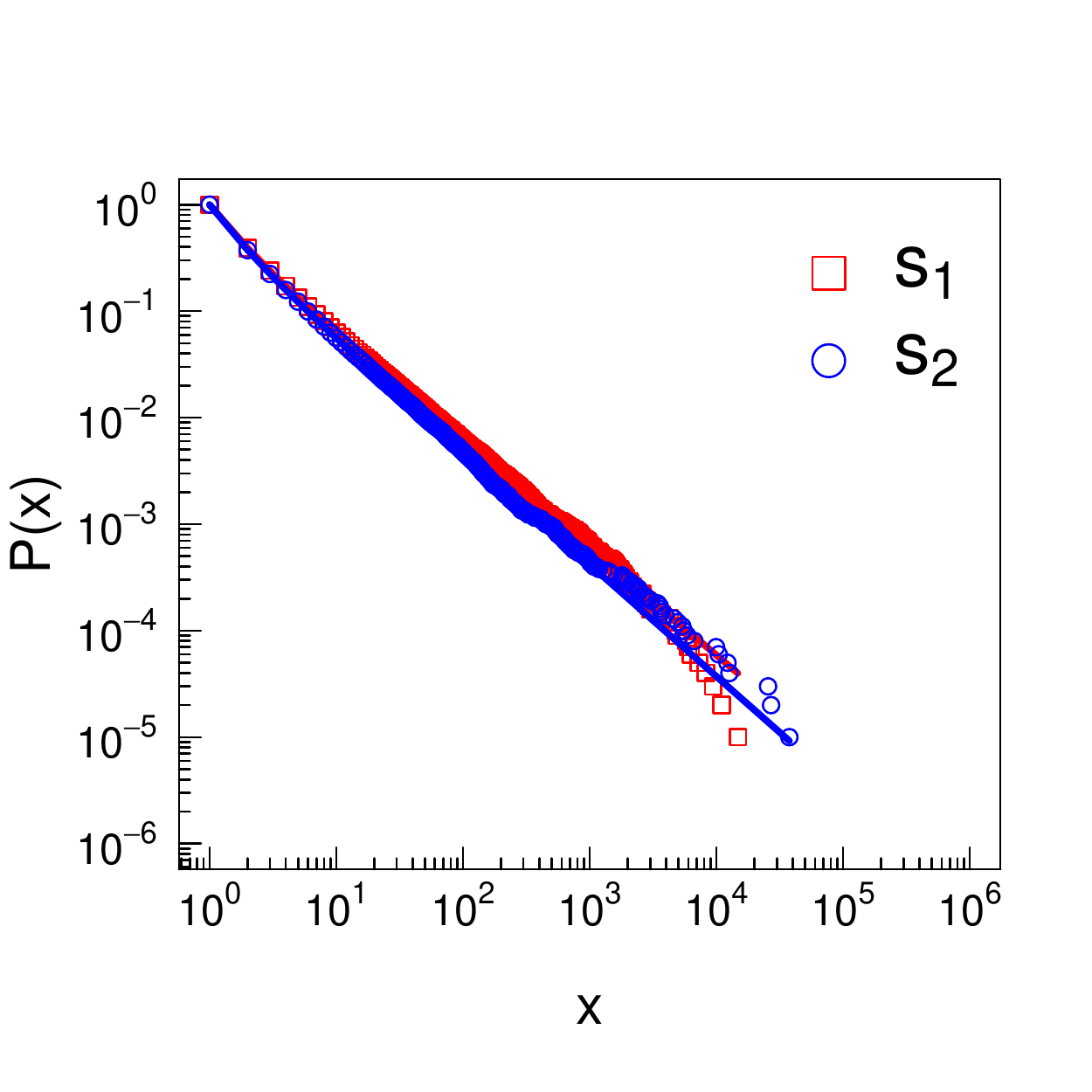}
	\caption{\textbf{Two samples drawn from different distributions.} Two samples of discrete data $s_{1}$ ($n_{s_{1}} = 10^{5}$) and $s_{2}$ ($n_{s_{2}} = 10^{5}$) are compared. The estimates of the scaling parameters are, respectively, $\hat{\alpha}_{s_{1}} = 2.003242$ and $\hat{\alpha}_{s_{2}} = 2.051032$. The test statistics is $\Lambda = 149.4912$ with p-value $< 10^{-3}$, thus we reject the null hypothesis and conclude that the two samples are not drawn from the same power-law distribution.}
	\label{fig:2} 
\end{figure}

\paragraph{Example 2: two samples drawn from different distributions.} Figure \ref{fig:2} shows the complementary cumulative distribution function, $P(x)$, of two samples, $s_{1}$ ($n_{s_{1}} = 10^{5}$) and $s_{2}$ ($n_{s_{2}} = 10^{5}$), that we want to compare in order to assess whether they are drawn from the same power-law distribution. The estimates of the scaling parameters are, respectively, $\hat{\alpha}_{s_{1}} = 2.003242$ and $\hat{\alpha}_{s_{2}} = 2.051032$. The test statistics is $\Lambda = 149.4912$ with p-value $< 10^{-3}$, thus we reject the null hypothesis and conclude that the two samples are not drawn from the same power-law distribution. Indeed, these samples were randomly drawn from two power-law distributions with different scaling parameters, respectively, $\alpha_{1} = 2$ and $\alpha_{2} = (2 + \delta)$, with $\delta = 0.05$.

\end{document}